**Author's approach to the topological modeling of parallel computing systems**

Victor A. Melent'ev

Rzhanov Institute of Semiconductor Physics, Siberian Branch of the Russian Academy of Sciences (ISP SB RAS)

Address: 13 Lavrentiev aven., 630090 Novosibirsk, Russian Federation

e-mail: melva@isp.nsc.ru, tel.: +7-923-246-78-28

ORCID: https://orcid.org/0000-0002-4855-9589

Web of Science Researcher ID: R-1610-2016

**Abstract**

The author's research of topologies of parallel computing systems and the tasks solved with them, including the corresponding tools of their modeling, is summarized in the present paper. The original topological model of such systems is presented based on the modified Amdahl law. It allowed formalizing the dependence of the necessary number of processors and the maximal distance between information-adjacent vertices in a graph on the directive values of acceleration or efficiency. The dependences of these values on the system interconnection topology and on the information graph of the parallel task are also formalized. The tools for a comparative evaluation of these dependences, topological criteria and the functions of scaling and fault-tolerant operation of parallel systems are based on the author's technique of projective description of graphs and the algorithms used in it.

**Introduction**

The way to exaflop computations consists in integrating hundreds of thousands of individual computing elements (nodes) into systems containing millions of computing cores. At that, the crucial factor here is the switching network (interconnect) adequacy to the maximal joint use of these cores in solving big problems [1-6] and its fault-tolerance [7-11].

The author' research is aimed at solving the fundamental problem of creating the formal methods of architectural projecting of parallel computing systems (CS) and organizing their fault-tolerant operation. The essence of the topological constituent of this issue is that although the CS parallelism potential dependence on the interconnection topologies used for its construction is commonly known, such dependences are not formally reflected. This makes one be content with *a priori* subjective and often contradictory expert estimations of topologies, the differences of which from an *a posteriori* test estimation of a final product may be unacceptable.

The issue of the analysis and synthesis of CS topologies is rather widely presented in scientific literature, and it is traditionally solved with the methods of graph theory [12-15]. At that, bijective correlations are established between a set of system modules and a set of graph vertices, as well as between a set of communication lines and a set of graph edges. However, neither in the theory of networks and systems, nor in its fundamental basis – the theory of graphs – the issue of topologies is actually researched with determinative methods, excluding the need to enumerate variants. This may be explained by the low level of traditional graph description forms which set just elementary relations of adjacency. That initially determines the non-polynomial character of searching the relations of higher orders, necessary for a CS research, and makes the researchers concentrate their efforts on developing not labor-intensive heuristic, stochastic or genetic methods and algorithms. As these methods are characterized by the indetermination in the operative management of systems functioning, which leads to unpredictable consequences, one can actually speak only about the probabilistic-optimal management, the reliability of which is determined by the sample size in stochastic algorithms or the cardinality of the initial population in genetic algorithms. Using undetermined algorithms in the operational management of large systems may lead to the destructive consequences caused by a delayed (irrelevant) reaction to the current changes in the system and inadequacy (unreliability) of this reaction.

Increasing the size and complexity of the system leads to the narrowing (up to disappearance) of the zone of mutual compliance with the conditions of relevance and reliability, while the probability of contingencies grows significantly [14, 16].

Thus, the problem solved in this research is due to 1) the absence of a topological model of parallel computations setting formal and completely unambiguous dependences of the potential parallelism of tasks on the CS topology; 2) the absence of the quantitative topological criteria of parallelism, reflecting these dependences, and determined methods for calculating them; 3) the non-polynomial (combinatorially "explosive") dependence (on the CS size and topology) of the computational complexity of determined CS methods adaptation to the tasks solved with it, which is, moreover, unacceptable in large-scale systems, where the multiplicity of failures grows with the number of constituting elementary computation modules, the number of possible combinations growing factorially. The results obtained by the author are substantiated, reviewed and summarized, for the first time, into a unified approach in the present work in order to allow the reader to get a holistic view of the formalized solution of the above problems directly related to the topologies and, hence, to the analytical modeling of a parallel system, taking into account the topologies used in them.

**1. Topological model of parallel computing systems.** The study of the conditionality of potential parallelism capabilities of systems by their interconnect topologies implies abstracting from the restrictions associated with the presence of scalar (non- parallelizable) fragments in the parallel algorithm, which are taken into account by the classical version of Amdahl law [16-19]. However, after including, in compliance with the mentioned law, the "scalarity" factor, one more factor impedes the linear scaling of the computing system productivity – that is, dependence on the inter-processor exchange. But this dependence is related not only to the system network topology, but also to its technological characteristics, such as reporting speed, latency, bandwidth, etc. Thus, one comes to the conclusion about the need to demarcate the topological and technological factors by their impact on the actual productivity of the system and on the maximal (under directive criteria of efficiency) order of subsystems so that it does not change the relative order of compared topologies when changing the topological characteristics of interconnection or changing the classes of the tasks solved and data processed.

The essence and novelty of the author's approach consist in creating such topological model of parallel computations that would set the formal correspondence (abstracted from network technologies used by interconnection) of the topology to the key property of a parallel system, i.e. its potential parallelism. The lower boundary of parallelism potential is viewed here as a generalized measure unit and is determined by the maximal rank of the parallel information-fully connected task with a limited reachability of information-adjacent task branches [20]. The substantiation of the acceptability of such abstracting and the possibility to analyze purely topological aspects of increasing the potential parallelism are given in [21].

To estimate the impact of CS topology on its parallelism, we abstract from applications, considering them to be unlimited-parallelizable and containing no scalar fragments. We introduce the following symbols: $W$ and $w$ – computation volumes, measured with time, when solving an arbitrary task on one processor and on $p$ processors of a computing system; $Q$ ($p = 1$) and $q = Q/p$ – data volumes, measured with information units (bytes) and to be exchanged, corresponding to the number $p$ of enabled processors. Thus, we suppose that:

1. The parallel algorithm of the task does not contain scalar (non-parallelizable) fragments and allows breaking information-connected parallel branches into an arbitrary number $p$: $1 \leq p \leq \infty$.

2. The total volume of computations $W$ and the volume $Q$ of the data to be exchanged after breaking the task into $p$ parallel branches do not depend on the number of processors $p$ and are distributed between them uniformly: $w = W/p$ и $q = Q/p$.

3. The data scaling in the task with coefficient $m$ increases the volume of computations $W$ and the volume $Q$ of the data to be exchanged $m$ times.

As is well known, the factors hindering the parallelism increase proportional to the system size are, first, practical impossibility of the fully connected topology even in relatively small systems and, second, delays in inter-processor exchanges. Therefore, we add the following to the list of the assumptions in our model:

4. All processors of the system are identical. Their total number $n$ is sufficient for implementing $p$ parallel branches ($n > p$), and the initial input data distribution over the processors enabled in the parallel application is not required.

5. The total volumes $W$ and $Q$ do not depend on the communication network topology and on the used network technology ($NT$ – Network Technology), and there are no limitations of the minimal volumes of $w$ and $q$.

6. The time spent $T_{ND}(p)$ for exchanges between information-adjacent processors are proportional to the distances $L(p)$ between the CS subgraph vertices corresponding to these processors and the delay function $t_{NT}(q)$, which depends on the *NT* used in the system: $T_{ND}(p) = L(p) \times t_{NT}(q)$; here index *ND* is the abbreviation of Network Delay.

7. The computing and communication elements of a CS allow concurrent operations, at $T_{ND}(p) > w$, and the time of actual delays, taking into account the concurrency, is determined by the difference $T_{ND}(p) - w > 0$.

8. The combination of topology and *NT* used in the computing system guarantees the absence of network collisions and delays due to them.

The sixth of the above properties of the proposed model is trivial in the practice of building and using communication networks: as a rule, the time of reporting between the most remote elements is estimated as the diameter of the corresponding graphs, and this characterizes the worst communication delay. It should be noted that an indispensable characteristic of function $t_{NT}(q)$ is that it is inversely related to *p* and directly - to *Q*: $p_1 < p_2 \Rightarrow t_{NT}(Q/p_1) > t_{NT}(Q/p_2)$ and $q_1 > q_2 \Rightarrow t_{NT}(q_1) > t_{NT}(q_2)$. Using technologies with a different relation of the $t_{NT}(q)$ function would contradict the main objective of paralleling, i.e. reaching the required speed of implementing users applications and their required reliability, which is related to the increased complexity of algorithms and/or processed data volumes. However, unfortunately, increasing the number of processors *p* by $k_p$ times, that leads to a proportional decrease of the exchanged data unit volume *q*, decreases the time $t_{NT}(q)$ of physically adjacent processors information interaction not at the same proportions, but at those depending on the *NT* used in the system, i.e. with a certain technological (intrinsic in the used network technology) coefficient

$$k_{NT} = k_p^{-1} \cdot t_{NT}(q) / t_{NT}(k_p^{-1} \cdot q).$$

Modifying the well-known Amdahl law with clauses 6 and 7 of our model taken into account, we will obtain acceleration $S_p$ reached by parallelizing the task into *p* branches

$$S_p = \frac{W}{w + (L_S(p) \cdot t_{NT}(q) - w)} = \frac{W}{L_S(p) \cdot t_{NT}(q)}.$$

Then the maximal distance $L_S(p) \geq 1$ of the acceleration $S_p$ for the preset number of processors *p* will be found from

$$L_S(p) = \frac{W}{S_p \cdot t_{NT}(q)}.$$

If, for example, the priority is not acceleration $S_p$, but efficiency $E_p = S_p/p$ of the *p* processors enabled for solving the problem, then

$$E_p = \frac{W}{p \cdot L_E(p) \cdot t_{NT}(q)},$$

and the maximal distance $L_E(p)$ for the preset efficiency value $E_p$ will be

$$L_E(p) = \frac{W}{p \cdot E_p \cdot t_{NT}(q)}.$$

It is necessary to understand the differences in the conditionality sources of distance *L* and the number of processors *p* in the (*W*, *Q*)-task: if the maximal distance $L(p)^1$ at the preset *p* is conditioned by the function $t_{NT}(Q/p)$, i. e. the network technology used in the CS, then the maximal number of processors $p(L)$, which can be enabled at the distance $L$ accepted by the task, is determined only topologically, i. e. it depends only on the graph used in the system. Thus, the correlation between the calculation volumes $W$ and information interactions $Q$ in the parallel $(W, Q)$-task and the operation speed of the $NT$ used in the system determine the interdependence between the number of operating processors $p$ and the maximal acceptable distance $L(p)$ between the information-adjacent vertices of the relevant subgraph in the CS graph, while the topology determines the possibility of implementing such a subgraph. Taking into account that the distances between the unweighed graph $G(V, E)$ vertices of are determined by the number of hops and may, therefore, be expressed only in integers, we determine the maximal distance between the information-adjacent processors as the integer part of $L(p)$ and call it reachability $\partial(p)$:

$$\partial(p) = \lfloor L(p) \rfloor, 1 \leq \partial(p).$$

---

[1] Here and further the indices of the used estimation criteria (acceleration, efficiency, etc.) are insignificant and omitted.

The obtained $\partial(p)$ value characterizes the requirement for the topology used in the CS, which is as follows: the successful (in relation to providing the criteria $S$ and/or $E$ prescribed to the solution of the problem with parameters $W$ and $Q$) parallelizing into $p$ processors is only possible when the system topology guarantees, at least, one embedding of the task information graph into the CS graph, under which the distances between the information-adjacent vertices do not exceed $\partial(p)$. Note the intuitively apparent direct dependence of the embedding success on the distance $L(p)$: the larger probability of the presence of subgraphs, complying with the directive values of *p* and $S_p$ in the CS graph, corresponds to

> the larger accepted distance conditioned by a more high-speed *NT*.

The problem of embedding the task information graph in the computing system graph is one of the most important ones in the theory of computing systems. Its solution is related to the problem of isomorphism identification in the theory of graphs and is of unfading interest for experts in this field [18, 19, 22, 23]. Within the topological model described here, the solution of the task embedding problem, based on the changing of the relations of CS graph vertices adjacency for the relations of their limited reachability, is presented in [20]for the first time. In [21, 24], the topological indicators for the scaling of parallel systems and tasks were first introduced. The problem of increasing the potential of tasks paralleling in a system without increasing the number of its processors, but through modifying the initial topology by complementing its relations of the adjacency between the processors is studied in [25], the basic means of addressing and routing being unchanged. The problems of analyzing topological fault-tolerance of a scalable computing system and providing its tolerance to faults of preset multiplicity are considered in [26-28]. Therein, the notion of topological adequacy of a computing system and the tasks solved with it are given, and the CS topological fault-tolerance conditions are analyzed; the criterion of topological fault-tolerance, directly connecting topology with a potential parallelism of the system under the preset multiplicity of accepted faults, is proposed; the interconnection between the functions of topological scaling and topological fault-tolerance of the systems is determined, and the conditionality of the minimum of topological fault-tolerance by the computing system graph girth is shown.

**2. Projective description of graphs and its use for the analysis of CS topologies.** The theoretical basis for the analysis, comparison and synthesis of the topologies of parallel computational systems, efficient in relation to the potential parallelism, is the graph description method proposed in [29], which, for the first time, gave formal grounds for the transition from stochastic and heuristic methods for the analysis and synthesis of topologies to analytical ones. It should be noted that using this method in other applications, for example, when developing biochemical problem-oriented databases [30], also proved its efficiency. The essence of the technique is in describing a graph through projections, and here we cite only the aspects necessary for understanding the presented material.

The projection $P(v_j)$ of graph $G(V, E)$ is a multilevel construction, at the zero level of which there is vertex $v_j \in V$ chosen as the angle one; the subset of the 1$^{st}$ level vertices $V_{1j} \subset V$ generated by it contains all vertices of its environment $\mathcal{N}(v_j)$, and the $i$-th level ($i \geq 1$) is the total of vertex subsets, each of which is generated by the vertex of ($i$ - 1)-th level and is the environment of this vertex without the vertices preceding it in this projection. Thus, the relation of "vertex precedence/subset generation", actually, simulates the relation of the adjacency of the preceding vertex to the vertices of the subset generated by it. The formal recording of these relations in the bracket description of two arbitrary neighboring levels of the graph projection is as follows:

$$v_{i1}^{V_{i+1,1}}, ..., v_{ij}^{V_{i+1,j}}.$$

Here the vertices $v_{i1}$ and $v_{ij}$ of one of the $i$-th level subsets precede and are adjacent to the vertices of the subsets $V_{i+1,1}$ and $V_{i+1,j}$ of the upper $(i + 1)$-th level generated by them. The technique of building the bracket (projective) graph descriptions and their properties are described in detail in [31-34] and summarized in [35, 36]. Therefore, here we discuss only several of their properties which are essential for this work.

The ordered set of vertices $W(v_{ij}) = (v_0, v_{10}, …, v_{ij})$, which is a simple chain from $v_0$ to $v_{ij}$ - the chain length is $\partial(v_0, v_{ij}) = i$ - corresponds to the vertex of the $v_{ij}$ $k$-level projection $P_k(v_0)$ built from the angle vertex $v_0$. In the general case, some (except for the angle one) vertices of projection $P_k(v_0)$ may be $m_{ij}$-fold: $0 \leq m_{ij} \leq \Sigma_i C_i - \Sigma_i |V_i|$, where $C_i$ is the number of elements of the $i$-th level of projection $P_k(v_0)$, and $V_i \subset V$ is the set of graph vertices represented by the $i$-th projection level. If $m_{ij}$ is not equal to one, it means that there is a correspondent number of simple chains from the angle vertex $v_0$ to vertex $v_{ij}$.

The number of the level $i$ in projection $P(v_0)$ determines the remoteness of vertices $V_i$ of this level from the angle vertex $v_0$, and also the fact that the level $k_e$, redetermining the set of vertices of all lower projection levels of graph $G(V, E)$ to $V$ for the first time, corresponds to the eccentricity $e(v_0)$ of the angle vertex $v_0$ in projection $P(v_0)$:

$$e(v_0) = k_e \mid \cup_{i=0}^{k_e-1} V_i \subset V, \; \cup_{i=0}^{k_e} V_i = V.$$

This condition is called vertex completeness of projection, but it is not always sufficient to determine all graph oriented edges. The projection $P_k(v_0)$ of graph $G(V, E)$ will be complete only if it determines all of its vertices and lines. Thus, the necessary conditions for the projection completeness can be written as follows:

$$\cup_{i=0}^{k} V_i = V \text{ and } \cup_{i=0}^{k} E_i = E,$$

where $E_i = \{e_{uv} \mid \text{ and } \in V_{i-1}, v \in V_i\}$ is the set of oriented edges incident to the vertex pairs of the neighboring projection levels. As one can see, the condition of vertex completeness of any graph projection is taken up by the condition of completeness of its oriented edges.

The above is explained by the example of a graph with its 2$^{nd}$ level projections shown in Fig. 1.

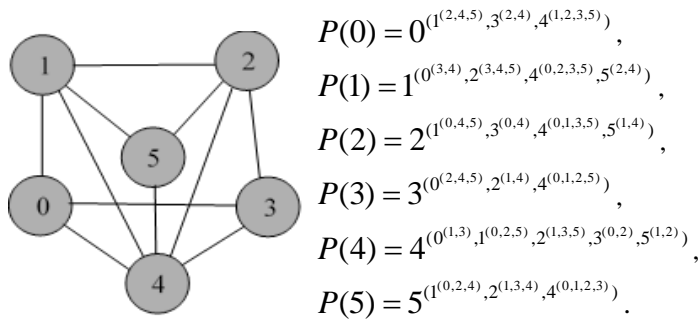

$P(0) = 0^{(1^{(2,4,5)}, 3^{(2,4)}, 4^{(1,2,3,5)})}$,

$P(1) = 1^{(0^{(3,4)}, 2^{(3,4,5)}, 4^{(0,2,3,5)}, 5^{(2,4)})}$,

$P(2) = 2^{(1^{(0,4,5)}, 3^{(0,4)}, 4^{(0,1,3,5)}, 5^{(1,4)})}$,

$P(3) = 3^{(0^{(2,4,5)}, 2^{(1,4)}, 4^{(0,1,2,5)})}$,

$P(4) = 4^{(0^{(1,3)}, 1^{(0,2,5)}, 2^{(1,3,5)}, 3^{(0,2)}, 5^{(1,2)})}$,

$P(5) = 5^{(1^{(0,2,4)}, 2^{(1,3,4)}, 4^{(0,1,2,3)})}$.

Fig. 1. Graph and its projections.

As any projection explicitly contains such metric characteristics, commonly used in the analysis of topologies, as the shortest and alternative paths, vertex eccentricities and graph diameter, then using the projective description of a graph delivers from the necessity of a rather complicated calculation of these characteristics, which is inevitable under the traditional description of graphs with adjacency/incidence matrices. Transforming the initial CS graph projections into the projections of a graph of preset reachability is also rather simple and is reduced to compressing initial projections [35, 36]. Thus, we significantly simplify the embedding of task information graph in the graph with the increased degree of vertices. In particular, for the tasks of ring structure, characteristic for signal and image processings [29], there is a description of the technique for revealing and enumerating the cycles with the length determined by the required acceleration.

In [30, 37], the problem of identifying a component in a computing system with a hypercubic topology that corresponds to a hosted subsystem, with respect to the reachability of its vertices, is considered. The formula of the maximal parallelizing of such subsystem was obtained, and the technique of specifying its elements was proposed. The idea of a possibility to use the graph projective description in order to create analytical methods of synthesizing systemic topologies with preset properties was first stated in [38, 39]. In [19, 24], the problem of synthesizing a computing system topology is solved as the problem of building a graph with a minimal diameter under the preset values of graph order, degree and girth. The solution is based on using the graph projective description and is reduced to building a unified (in relation to the above characteristics) system of its vertex-complete projections. Therein, the notion of a compact graph is introduced, its analytical model is determined, and the generation algorithm based on it is described.

As the main parameter of the approach described in the present paper, which characterizes the quality of topology, is the potential parallelism of fully connected information tasks provided by it, below we demonstrate the use of graphs projective description for revealing the fully connected subgraphs in the graph shown in Fig. 1 – a clique of this graph. The algorithm of revealing cliques and its demonstration on more complex examples, including reachability values $\partial \geq 1$ and the preset multiplicity values, the accepted failures are described in more detail in [40].

The first level of graph projection $P(0)$ (Fig. 1) contains 3 vertices: $V_1(0) = \{1,3,4\}$. It means that the order of the maximal clique $K(G)$ containing vertex 0 cannot be more than four ($|K(G)| \leq 3 + 1$). Taking into account that the 2$^{nd}$ level vertex subsets cannot include the vertices not belonging to $V_1(0)$, we exclude such vertices from the projection, leaving, in the 2$^{nd}$ level subsets, only the 1$^{st}$ level vertices:

$P(0) = 0^{(1^{(4)}, 3^{(4)}, 4^{(1,3)})}$.

The maximal 2nd level subset– $M(0) = (1,3)$ – is the only one and contains only two ($m(0) = |M(0)| = 2$) vertices; if such subset were not the only one but as many as the 1st level vertices $n_1(0) = 3$, then we could consider the possibility for the existence of a clique of the 4th order. However, the subset with cardinality $m(0) = 2$ is the only one; thus, the maximal clique cannot have the order of more than three. It is easy to see such cliques (their vertices are written in bold here):

$P(0) = \mathbf{0}^{(\mathbf{1}^{(4)},3^{(4)},\mathbf{4}^{(1,3)})}$, clique (0, 1, 4),

$P(0) = \mathbf{0}^{(1^{(4)},\mathbf{3}^{(4)},\mathbf{4}^{(1,3)})}$, clique (0, 3, 4).

But as we need not any, but the most maximal clique, then we do the same for the other five graph projections, just as for $P(0)$:

$P(1) = 1^{(0^{(4)},2^{(4,5)},4^{(0,2,5)},5^{(2,4)})}$,

$P(2) = 2^{(1^{(4,5)},3^{(4)},4^{(1,3,5)},5^{(1,4)})}$,

$P(3) = 3^{(0^{(2,4)},2^{(4)},4^{(0,2)})}$,

$P(4) = 4^{(0^{(1,3)},1^{(0,2,5)},2^{(1,3,5)},3^{(0,2)},5^{(1,2)})}$,

$P(5) = 5^{(1^{(2,4)},2^{(1,4)},4^{(1,2)})}$.

As we can see, projection $P(3)$, just as $P(0)$, does not comply with the requirement of the order of more than 3 for clique ($m(3) > 2$). Therefore, we consider and correct the other 4 projections, excluding vertices 0 and 3 from the subsets of these projections (above we showed that these vertices cannot be within the cliques having orders of more than three):

$$P(1) = 1^{(2^{(4,5)}, 4^{(2,5)}, 5^{(2,4)})},$$

$$P(2) = 2^{(1^{(4,5)}, 4^{(1,5)}, 5^{(1,4)})},$$

$$P(4) = 4^{(1^{(2,5)}, 2^{(1,5)}, 5^{(1,2)})},$$

$$P(5) = 5^{(1^{(2,4)}, 2^{(1,4)}, 4^{(1,2)})}.$$

In each of the obtained projections, the subsets formed by any 1st level vertex and the 2nd level vertices generated by it coincide; thus, the largest clique in the studied graph consists of four vertices (1, 2, 4, 5). Consequently, the potential parallelism of such CS graph, when solving the fully connected information problems, is $\varphi_{\partial=1}(G) = 4$.

**Conclusion**

During their exploitation, all parallel systems are researched for their operating speed and efficiency by various classes and sets of tasks, and data. However, the differences in technical, technological, topological, applied and other architectural components of the systems make the obtained results exclusive and can be extended to other systems and tasks only with certain limitations. That hinders obtaining the integral picture of the formal conditionality of parallelism on the systems topology. An attempt to bridge the gap existing in this field was made in this research. To this end, we propose the parallel calculations model divided into two constituents: the first one refers to parallel applications and attributes the properties of unlimited paralleling to them; the second one refers to the computing system in which the parallelism limitations are conditioned by the maximally allowed distance between the information-adjacent CS graph vertices, connected with the operating speed of its communication environment and the volumes of computation, and exchange operations of the parallel task.

Using the proposed model allowed focusing on the technological and topological constituents of the communication environment and researching its formal interrelation in the aggregate impact on the limits of paralleling the tasks corresponding to the preset efficiency criteria. We obtained the formal expressions connecting the maximally allowed distances between the information-adjacent processors $L(p)$ and the number of parallel branches with preset acceleration values.

The research, optimization and synthesis of topologies of computing systems interconnection are proposed to be based on the projective description of graphs operating with the relations of the order higher than the relations of vertices adjacency. The author describes the method and its application to graphs with limited distances (reachability) between information-adjacent vertices. The problem of embedding parallel tasks into a CS graph was considered, taking into account the issue of limited reachability. Such notions as graph of $\partial$-reachability, $\partial$-components of a graph being a clique of the graph of $\partial$-reachability, $\partial$-density of CS graph and the algorithm of revealing such cliques in a CS graph were introduced. The mutually abstracted formalized indicators and functions of the topological scaling of parallel tasks and topological scaling of the systems enable, without being tied to the specific environment of the task implementation, choosing the least topologically complex algorithm for embedding or choosing the topologies which - other conditions being equal - show the largest possibilities for a successful embedding of a set of certain tasks.

The research results will be useful not only for analyzing the existing systems and parallel algorithms implemented in them, but also for creating new systems or algorithms, taking into account the supposed scaling of both of them.

**Acknowledgements**

The author would like to express his sincere gratitude to the Russian Foundation for Basic Research for the support of the present research with grants 98-01-00402, 01-01-00790, 05-08-01301 and 14-07-00169 during several years.

**References**

[1] Wu A. Y. Embedding of tree networks into hypercubes, Journal of Parallel and Distributed Computing, Volume 2, Issue 3 (1985) 238–249. ISSN 0743-7315, https://doi.org/10.1016/0743-7315(85)90026-7 (accessed 20 June 2019).

[2] Hayter, Th., Brookes, G. R. Approach to the simulation of various LAN topologies, Computer Communications, Volume 12, Issue 4 (1989) 204–212.

ISSN 0140-3664, https://doi.org/10.1016/0140-3664(89)90197-7 (accessed 20 June 2019).

[3] Caselli, S., Conte, G., Malavolta, U. Topology and process interaction in concurrent architectures: A GSPN modeling approach, Journal of Parallel and Distributed Computing, Volume 15, Issue 3 (1992) 270–281. ISSN 0743-7315. https://doi.org/10.1016/0743-7315(92)90008-B (accessed 20 June 2019).

[4] Lloret, J., Palau, C., Boronat, F., Tomas, J. Improving networks using group-based topologies, Computer Communications, Volume 31, Issue 14 (2008) 3438–3450. ISSN 0140-3664. https://doi.org/10.1016/j.comcom.2008.05.030 (accessed 20 June 2019).

[5] Abd-El-Barr, M., Gebali, F. Reliability analysis and fault tolerance for hypercube multi-computer networks, Information Sciences, Volume 276 (2014) 295–318. ISSN 0020-0255. https://doi.org/10.1016/j.ins.2013.10.031 (accessed 20 June 2019).

[6] Emmert-Streib, F., Dehmer, M.,Yongtang Shi. Fifty years of graph matching, network alignment and network comparison, Information Sciences, Volumes 346–347 (2016) 180–197. ISSN 0020-0255. https://doi.org/10.1016/j.ins.2016.01.074 (accessed 20 June 2019).

[7] Das, Ch. R., Bhuyan, L. N. Dependability evaluation of interconnection networks, Information Sciences, Volume 43, Issues 1–2 (1987) 107–138. ISSN 0020-0255. https://doi.org/10.1016/0020-0255(87)90033-8 (accessed 20 June 2019).

[8] Ke Huang, Jie Wu. Fault-tolerant resource placement in balanced hypercubes, Information Sciences, Volume 99, Issues 3–4 (1997) 159–172. ISSN 0020-0255. https://doi.org/10.1016/S0020-0255(96)00270-8 (accessed 20 June 2019).

[9] Jianxi Fan, Xiaohua Jia, Xin Liu, Shukui Zhang, Jia Yu. Efficient unicast in bijective connection networks with the restricted faulty node set, Information Sciences, Volume 181, Issue 11 (2011) 2303–2315. ISSN 0020-0255. https://doi.org/10.1016/j.ins.2010.12.011 (accessed 20 June 2019).

[10] Chechina, N., Huiqing Li, Ghaffari, A., Thompson, S., Trinder, Ph. Improving the network scalability of Erlang, Journal of Parallel and Distributed Computing, Volumes 90–91 (2016). 22–34. ISSN 0743-7315. https://doi.org/10.1016/j.jpdc.2016.01.002 (accessed 20 June 2019).

[11] Hutchison, D., Sterbenz, J. P.G. Architecture and design for resilient networked systems, Computer Communications, Volume 131 (2018) 13–21. ISSN 0140-3664. https://doi.org/10.1016/j.comcom.2018.07.028 (accessed 20 June 2019).

[12] Levi, G., Luccio, F. A technique for graph embedding with constraints on node and arc correspondences, Information Sciences, Volume 5 (1973) 1-24. ISSN 0020-0255. https://doi.org/10.1016/0020-0255(73)90001-7 (accessed 20 June 2019).

[13] Grout, V. M., Sanders, P. W. Communication network optimization, Computer Communications, Volume 11, Issue 5 (1988) 281–287. ISSN 0140-3664. https://doi.org/10.1016/0140-3664(88)90039-4 (accessed 20 June 2019).

[14] Dutt, Sh., Hayes, J. P. Designing fault-tolerant systems using automorphisms, Journal of Parallel and Distributed Computing, Volume 12, Issue 3 (1991) 249–268. ISSN 0743-7315. https://doi.org/10.1016/0743-7315(91)90129-W (accessed 20 June 2019).

[15] Jianxi Fan, Xiaohua Jia. Edge-pancyclicity and path-embeddability of bijective connection graphs, Information Sciences, Volume 178, Issue 2 (2008) 340–351. ISSN 0020-0255. https://doi.org/10.1016/j.ins.2007.08.012 (accessed 20 June 2019).

[16] Subharthi, P., Jianli Pan, Raj Jain. Architectures for the future networks and the next generation Internet: A survey, Computer Communications, Volume 34, Issue 1 (2011) 2–42. ISSN 0140-3664. https://doi.org/10.1016/j.comcom.2010.08.001 (accessed 20 June 2019).

[17] Xian-He Sun, Yong Chen. Reevaluating Amdahl's law in the multicore era, Journal of Parallel and Distributed Computing, Volume 70, Issue 2 (2010) 183–188. ISSN 0743-7315. https://doi.org/10.1016/j.jpdc.2009.05.002 (accessed 20 June 2019).

[18] Hao Che, Minh Nguyen. Amdahl's law for multithreaded multicore processors, Journal of Parallel and Distributed Computing, Volume 74, Issue 10

(2014) 3056–3069. ISSN 0743-7315. https://doi.org/10.1016/j.jpdc.2014.06.012 (accessed 20 June 2019).

[19] Yavits, L., Morad, A., Ginosar, R. The effect of communication and synchronization on Amdahl's law in multicore systems, Parallel Computing, Volume 40, Issue 1 (2014) 1–16. ISSN 0167-8191. https://doi.org/10.1016/j.parco.2013.11.001 (accessed 20 June 2019).

[20] Melent'ev, V. A. Embedding of subsystems limiting length and number of paths between vertexes of computing system graph, UBS, 47 (2014) 212–246. http://mi.mathnet.ru/eng/ubs749 (accessed 20 June 2019).

[21] Melent'ev, V. A. On topological scalability of computing systems, UBS, 58 (2015) 115–143. http://mi.mathnet.ru/eng/ubs844 (accessed 20 June 2019).

[22] Xian-He Sun, Yong Chen. Reevaluating Amdahl's law in the multicore era, Journal of Parallel and Distributed Computing, Volume 70, Issue 2 (2010) 183–188. ISSN 0743-7315. https://doi.org/10.1016/j.jpdc.2009.05.002 (accessed 20 June 2019).

[23] Melent'ev ,V.A., Shubin, V.I., Zadorozhny, A.F. Topological scalability of hypercubic parallel systems and tasks. ISJ Theoretical & Applied Science 11 (31) (2015) 122–129. Doi: http://dx.doi.org/10.15863/TAS.2015.11.31.19 (accessed 20 June 2019).

[24] Nutaro, J., Zeigler, B. How to apply Amdahl's law to multithreaded multicore processors, Journal of Parallel and Distributed Computing, Volume

107 (2017) 1–2. ISSN 0743-7315. https://doi.org/10.1016/j.jpdc.2017.03.006 (accessed 20 June 2019).

[25] Melentiev, V. Edge scaling of computing systems, UBS, 64 (2016) 81–11. http://mi.mathnet.ru/eng/ubs898 (accessed 20 June 2019).

[26] Melent'ev, V. A. On topological fault-tolerance of scalable computing systems, UBS, 70 (2017) 58–86. URL: https://doi.org/10.25728/ubs.2017.70.3 (accessed 20 June 2019).

[27] Melent'ev, V. A. Fault-tolerance of hypercubic and compact topology of computing systems. ISJ Theoretical & Applied Science, 12 (44) (2016) 98–105. Doi: http://dx.doi.org/10.15863/TAS.2016.12.44.20 (accessed 20 June 2019).

[28] Melent'ev, V. A. On approach to the configuring of fault-tolerant subsystems in case of scarce topological fault-tolerance of the computing system. ISJ Theoretical & Applied Science, 10 (54) (2017) 101–105. Doi: https://dx.doi.org/10.15863/TAS.2017.10.54.20 (accessed 20 June 2019).

[29] Melent'ev, V. A. Bracket form of the graph description and its use in the structural investigations of enduring computer systems, Avtometriya, 4 (2000) 36–51. https://elibrary.ru/item.asp?id=14954075 (accessed 20 June 2019).

[30] Kornushk, V. F., Bogunova, I. V., Flid, A. A., Nikolaeva, O. M., Grebenshchikov, A. A. Information-algorithmic support for development of

solid pharmaceutical form. Fine Chemical Technologies, 13(5) (2018) 73–81. https://doi.org/10.32362/2410-6593-2018-13-5-73-81 (accessed 20 June 2019).

[31[ Melentiev, V. A. The bracket pattern of a graph. 6[th] International Conference on Pattern Recognition and Image Analysis: New Information Technologies, PRIA-6-2002, October 21-26 (2002) Proceedings, Velikiy Novgorod, Russian Federation, 57–61.

[32] Melentiev, V. A. Formalnye osnovy skobochnyh obrazov v teorii grafov [Formal bases of bracket images in graph theory]. 2[nd] International Conference "Parallel computations andy management tasks" PACO'2004: In-t problem upravleniya RAN im. V.A. Trapeznikova, (2004) 694–706.

[33] Melentiev, V. A. Formalnyy podkhod k issledovaniyu struktur vychislitelnykh system [Formal approach to researching the structures of computation systems]. Vestnik Tomskogo gosudarstvennogo universiteta,14 (2005) 167–172.

[34] Melent'ev, V. A. The metric, cyclomatic and synthesis of topology of systems and networks (2012). https://elibrary.ru/download/elibrary_22249411_42044045.pdf (accessed 20 June 2019).

[35] Melent'ev, V. A. Use of Melentiev's graph representation method for identification and enumeration of circuits of the given length. ISJ Theoretical & Applied Science, 11 (67) (2018) 85–91. Doi: https://dx.doi.org/10.15863/TAS.2018.11.67.16 (accessed 20 June 2019).

[36] Melentiev, V. A., Limit configuring of subsystems in hypercubic computing systems, Journal of Information Technologies and Computing Systems, 2 (2015) 20–30. http://www.jitcs.ru/images/documents/2015-02/20_30.pdf (accessed 20 June 2019).

[37] Melentiev, V. A. The limiting paralleling in the computing system with a hypercube topology under length restriction of interprocess connections. https://elibrary.ru/item.asp?id=23316608 (accessed 20 June 2019).

[38] Melent'ev, V. A. An analytical approach to the synthesis of regular graphs with preset values of the order, degree and girth, Prikl. Diskr. Mat., 2(8), (2010) 74–86. http://mi.mathnet.ru/eng/pdm178 (accessed 20 June 2019).

[39] Volkova, A. A Technical Translation of Melentiev's Graph Representation Method with Commentary. University Honors Theses. Paper 503 (2018). http://dx.doi.org/10.15760/honors.507 (accessed 20 June 2019).

[40] Melent'ev, V. A. Use of Melentiev's graph representation method for detection of cliques and the analysis of topologies of computing systems. ISJ Theoretical & Applied Science, 12 (68), (2018) 201–211. Doi: https://dx.doi.org/10.15863/TAS.2018.12.68.28 (accessed 20 June 2019).